\documentclass[aps,prl,twocolumn,superscriptaddress,amsmath,amssymb]{revtex4-2}
\usepackage{hyperref,orcidlink}
\usepackage[T1]{fontenc}
\usepackage[utf8]{inputenc}
\usepackage{lmodern}
\hypersetup{colorlinks=true, citecolor=blue, urlcolor=blue, linkcolor=blue} 
\renewcommand{\section}[1]{{\par\it #1.---}\ignorespaces}

\begin{document}
\title{Quantum gravimetry with mechanical qubits}
\author{Xiao-Wen Huo}
\affiliation{Ministry of Education Key Laboratory for Nonequilibrium Synthesis and Modulation of Condensed Matter, Shaanxi Province Key Laboratory of Quantum Information and Quantum Optoelectronic Devices, School of Physics, Xi'an Jiaotong University, Xi'an 710049, China}
\author{Jun-Hong An}
\affiliation{Key Laboratory of Quantum Theory and Applications of MoE, Lanzhou Center for Theoretical Physics, and Key Laboratory of Theoretical Physics of Gansu Province, Lanzhou University, Lanzhou 730000, China}
\author{Peng-Bo Li}
\email{lipengbo@mail.xjtu.edu.cn}
\affiliation{Ministry of Education Key Laboratory for Nonequilibrium Synthesis and Modulation of Condensed Matter, Shaanxi Province Key Laboratory of Quantum Information and Quantum Optoelectronic Devices, School of Physics, Xi'an Jiaotong University, Xi'an 710049, China}

\begin{abstract}
Levitated mesoscopic particles hold the promise of revolutionizing gravity sensing by using quantum effects. However, conventional quantum gravimeters based on such systems fail to harness the intrinsic large-mass advantage of the particles, because their commonly utilized auxiliary quantum systems counteract the role of mass as a resource. To overcome this limitation, we propose a quantum gravimetry by directly using the mechanical qubit (QM) formed by a levitated particle as the gravity sensor. Without resorting to the auxiliary quantum system, our scheme enables a straightforward readout of the particle's motion under gravitational influence. The obtained sensitivity behaves as a $m^{-1/2}$-scaling with the mass $m$. We also generalize our scheme to the \textit{mechanical cat qubit} as the gravity sensor. The sensitivity further scales as $N^{-1/2}$ with the mean phonon number $N$. In the experimentally realizable parameter regime, a sensitivity on the order of $0.1~ \text{\textmu}\text{Gal}/\sqrt{\text{Hz}}$ can be achieved, which outperforms the traditional schemes by two orders of magnitude. Reaching the \textit{double standard quantum limits} with $m$ and $N$ simultaneously, our scheme provides a feasible route toward compact high-sensitivity quantum gravimetry.
\end{abstract}
\maketitle

\section{Introduction} 
High-precision gravimeters play a vital role in various domains, ranging from basic scientific discovery~\cite{2004JentschP21972221,doi:10.1126/sciadv.adk2949,PhysRevLett.100.031101,Belenchia2022P170} to engineering applications such as geological surveying~\cite{2025CrawfordP861876,2021ForsterP2626}, as well as military operations, including underwater and underground mappings~\cite{2022StrayP590594,Krelina2021P2424}. Taking advantage of the extreme sensitivity of quantum systems to external perturbations~\cite{2013MoserP493496,2021FoglianoP41244124,https://doi.org/10.1002/adma.202109621,10.1063/1.4881936,LIANG202357,2025TsengP4036740367,2026MiP6670166701}, quantum metrology enables the realization of high-precision measurements of weak signals, advancing both fundamental science and frontier technologies~\cite{Montenegro_2025,Jiao2025P23002182300218,qk5r-h851,6gql-zgkb,2026GrochowskiP,2026LiP374378,2026ZhangP18251825,2026ZhouP1031410314,2024XuP14721477,2022BarbieriP1020210202,Jing:18,2018PezzeP3500535005,2014TothP424006424006,2011GiovannettiP222229,2006GiovannettiP1040110401}.
In this context, quantum gravimetry~\cite{2025WeiP1206412064,PhysRevLett.132.190001,2022ZhongP100230100230,2020NobiliP1203612036,2019GietkaP190801190801,2011PoliP3850138501} represents a significant technological advance. It enables extraordinary sensitivity in measuring gravity by harnessing the principles of quantum mechanics. 
Levitated mesoscopic particles have garnered significant attention as a promising platform for high-precision gravimeters~\cite{PhysRevLett.111.180403,2018QvarfortP36903690,PhysRevLett.121.063602,2024LengP123601123601,z8b4-sm79,RevModPhys.97.015003,RademacherMillenLi+2020+227+239}. This interest comes from two key advantages. Firstly, the particles are free from mechanical constraints. Secondly, their gravitational response scales as $\sqrt{m}$, providing a pronounced susceptibility to gravity due to their large mass $m$~\cite{z8b4-sm79,RademacherMillenLi+2020+227+239,RevModPhys.97.015003,doi:10.1126/science.abg3027}.

In conventional gravity-sensing schemes with levitated particles, the minute displacements induced by gravity are difficult to measure directly due to the harmonic nature of the mechanical motion. This has led to the need for coupling to an auxiliary spin qubit ~\cite{PhysRevLett.111.180403,z8b4-sm79,RevModPhys.97.015003} or a cavity mode~\cite{2021TebbenjohannsP378382,RevModPhys.97.015003,2018QvarfortP36903690}.
However, the introduction of an auxiliary two-level system (TLS) or a cavity leads to a mass-independent sensitivity, because the coupling strength between the particle and the auxiliary system scales as $1/\sqrt{m}$~\cite{PhysRevLett.125.153602,2016LiP1550215502,z8b4-sm79,PhysRevLett.124.163604}, undesirably canceling the $\sqrt{m}$ enhancement from the particle's mass. Under these circumstances, conventional gravimetry schemes based on levitated mesoscopic particles fail to exploit the $\sqrt{m}$ enhancement provided by the particle's large mass~\cite{2018QvarfortP36903690,z8b4-sm79,RevModPhys.97.015003,PhysRevLett.111.180403}, thereby limiting the performance of such gravimeters. To overcome this limitation, a critical step is to identify a quantum system that can realize gravimetry by directly exploiting the levitated particles, without relying on auxiliary sensors. The mechanical qubit (MQ), which combines the quantum properties of mechanical modes with a two-level structure, is a promising candidate for this role. Since its proposal in 2013~\cite{2013RipsP120503120503}, researchers have actively pursued its realization by leveraging Duffing nonlinearity to enhance the anharmonicity of mechanical modes~\cite{3mlc-r8x4,PhysRevX.11.031027,2023SamantaP13401344,sharma2025micromechanicalqubitbasedquantized}, with several experimental demonstrations now reported~\cite{2024YangP783788}. The development of MQs thus marks a significant advance in the extension of mechanical systems to universal quantum computing and precision measurement~\cite{3mlc-r8x4,doi:10.1126/science.adt2497}.

Here, we propose a mass-enhanced quantum gravimetry based on a MQ encoded in the center-of-mass (CM) mode of a levitated particle with Duffing nonlinearities. This MQ can  directly respond to gravity without coupling to an auxiliary system. Consequently, our scheme can fully leverage the advantage of the particle's large mass, providing a fundamental advantage over previous schemes that rely on coupling to an auxiliary TLS or cavity. We show that the sensitivity for gravity sensing scales as $m^{-1/2}$ with the mass $m$. Moreover, we also present an advanced quantum gravimetry with mechanical cat qubits (MCQs) by integrating MQs with mechanical cat states~\cite{PhysRevLett.130.213604}. This scheme further enhances the sensitivity by a factor of $N^{-1/2}$ over the MQ gravimetry scheme, where $N$ denotes the mean phonon number of the cat state, and thereby achieves the double standard quantum limits~\cite{1981CavesP16931708,RevModPhys.89.035002,2021WuP5404254042,2019LiuP2300123001}, in both the $N$ and $m$ scalings. The obtained sensitivity reaches $0.1~ \text{\textmu}\text{Gal}/\sqrt{\text{Hz}}$, which is smaller than the ones in the traditional single-particle schemes by two orders of magnitude~\cite{PhysRevLett.111.180403,z8b4-sm79,2024LengP123601123601}. Given that mechanical cat states have been realized~\cite{doi:10.1126/science.adf7553}, our quantum gravimeter is experimentally feasible. In particular, our scheme employs a Rabi-measurement approach~\cite{RevModPhys.89.035002}, which eliminates the need for Ramsey interference and free falling processes in conventional schemes~\cite{z8b4-sm79,PhysRevLett.121.063602,PhysRevLett.111.180403}. This leads to a significantly compact quantum gravimeter that is straightforward to implement.

\section{The MQ gravimeter}
\begin{figure}
\centering
\includegraphics[width=0.48\textwidth]{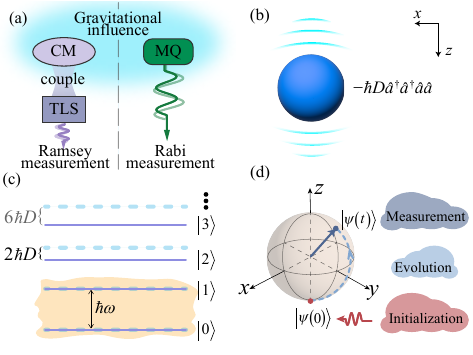}
\caption{(a) Schematic comparison of our quantum MQ gravimetry with the previous TLS-assisted one. The blue background denotes the influence of gravity. (b) Model used to construct the MQ gravimeter. (c) Energy spectra of the system with $D\neq 0$ in the purple solid lines and with $D=0$ in the blue dashed lines. (d) Rabi measurement in the MQ gravimetry and its representation on the Bloch sphere.}\label{FIG1}
\end{figure}

A quantum gravimeter typically employs a levitated mesoscopic particle, whose CM motion serves as a mechanical oscillator, as the gravitational probe~\cite{z8b4-sm79,RevModPhys.97.015003,RademacherMillenLi+2020+227+239}. In conventional schemes, an auxiliary TLS is introduced to couple to the CM mode, enabling the gravitational signal to be encoded in its internal-state populations for readout~\cite{z8b4-sm79,RevModPhys.97.015003}. However, the introduction of a TLS makes the detection outcome independent of the particle mass~\cite{2018QvarfortP36903690,z8b4-sm79,RevModPhys.97.015003,PhysRevLett.111.180403}. In our gravimeter, the MQ realized in the CM mode of a levitated particle serves directly as the sensor, simultaneously sensing the gravitational acceleration and encoding the resulting signal, thereby eliminating the need for an auxiliary TLS. This approach retains the mass-dependent sensitivity enhancement. Figure~\ref{FIG1}(a) schematically compares the conventional approach with our method.

Firstly, we propose to use a levitated mesoscopic particle whose CM mode exhibits Duffing nonlinearity as the quantum gravimeter, see Fig.~\ref{FIG1}(b). Its free Hamiltonian reads $\hat{H}_{1}=\hbar\omega\hat{a}^{\dagger}\hat{a}-\hbar D\hat{a}^{\dagger}\hat{a}^{\dagger}\hat{a}\hat{a}$, where $\hat{a}$ is the annihilation operator of the CM mode with frequency $\omega$ along the gravitational direction and $D$ characterizes the strength of the Duffing nonlinearity. The eigenenergies of the system are $E_n=n\hbar\omega-n(n-1)\hbar D$, which show an anharmonicity $\Delta_{1}=2\hbar D$ between the first and second energy spacings~\cite{2024YangP783788,2019KrantzP2131821318}, see Fig.~\ref{FIG1}(c). Under the condition that external perturbations are much smaller than the anharmonicity $\Delta_{1}$, the dynamics of the system is restricted to the subspace spanned by the ground state $|0\rangle$ and the first excited state $|1\rangle$. This allows for the realization of a MQ with Pauli operators $\hat{\sigma}_{1}^{z}=|1\rangle\langle1|-|0\rangle\langle0|$ and $\hat{\sigma}_{1}^{x}=|1\rangle\langle0|+|0\rangle\langle1|$~\cite{2024YangP783788,2019KrantzP2131821318,PhysRevX.11.031027}. In the presence of the gravitational effect, the Hamiltonian of the quantum gravimeter becomes
$\hat{H}^{g}_{1}=\hat{H}_1+(mg-F)z_{0}(\hat{a}^{\dagger}+\hat{a})$,
where $g$ is the gravitational acceleration and $z_{0}=\sqrt{\hbar/(2m\omega )}$ is the zero-point fluctuation amplitude of the CM motion. Here, $F$ is an applied static force along the direction opposite to gravity to ensure $(mg-F)z_{0}\ll\Delta_{1}$, so that the system remains confined to the computational subspace. Projecting $\hat{H}^{g}_{1}$ into the MQ subspace yields
\begin{equation}
\hat{\mathcal H}^{g}_{1}=\hbar\frac{\omega}{2}\hat{\sigma}_{1}^{z}+\hbar\Omega_{1}\hat{\sigma}_{1}^{x}\label{H1g'}
\end{equation}
with $\Omega_{1}\equiv(mg-F)z_{0}/\hbar$. Our quantum gravimetry is based on the Rabi measurement scheme~\cite{RevModPhys.89.035002}. It consists of the following steps, see Fig.~\ref{FIG1}(d). First, the state of the MQ is initialized in the ground state $|\psi_{1}(0)\rangle=|0\rangle$. Then, the dynamics governed by Eq. \eqref{H1g'} encodes the gravitational acceleration in the state of the MQ as $|\psi_{1}(t)\rangle=\left[\cos{(\mathcal{A}t/2)}+i\omega/\mathcal{A}\sin{(\mathcal{A}t/2)}\right]|0\rangle-2i\Omega_{1}/\mathcal{A}\sin{(\mathcal{A}t/2)}|1\rangle$ with $\mathcal{A}\equiv\sqrt{\omega^{2}+4\Omega_{1}^{2}}$.
Third, a projective measurement to the excited-state occupation operator $\hat{O}_{1}=|1\rangle\langle 1|$ gives the mean value as $\bar{O}_{1}=4\Omega_{1}^{2}/\mathcal{A}^{2}\sin^{2}{(\mathcal{A}t/2)}$~\cite{SM}.
 
\begin{figure}
\centering
\includegraphics[width=0.48\textwidth]{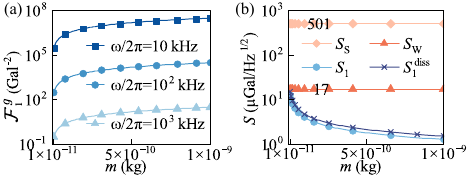}
\caption{(a) Mass-dependent variation of the QFI in the MQ gravimetry scheme with different frequency $\omega$. (b) Comparison of our gravitational sensitivities $S_1$ and $S_1^{\rm{diss}}$ with $S_\text{S}$ in Ref.~\cite{PhysRevLett.111.180403} and $S_\text{W}$ in Ref.~\cite{z8b4-sm79} as a function of the particle mass $m$ with $\omega/2\pi= 10~\text{kHz}$.  Other parameters to calculate $S_1^{\rm{diss}}$ are $Q=10^8$, $T=10~\rm{mK}$ and $\Omega_1=0.02\omega$. The vertical axes in both panels are logarithmic.}	\label{FIG2}
\end{figure}

The ultimate precision of any quantum sensor to detect a quantity $\theta$ from the sensor state $\rho_\theta$ is constrained by the quantum Cram\'{e}r-Rao bound $\delta\theta= (\upsilon \mathcal{F}^{\theta})^{-1/2}$, where $\delta\theta$ is the root-mean-square error of $\theta$, $\upsilon$ is the number of repeated experiments, and $\mathcal{F}^{\theta}\equiv \mathrm{Tr}(\hat{\varsigma}^{2}\rho_{\theta})$ is the so-called quantum Fisher information (QFI) ~\cite{DeMille2024P741749,Riedel2010P11701173,2018QvarfortP36903690}. $\hat{\varsigma}$ is the symmetric logarithmic derivative and determined by $\partial_{\theta}\rho_{\theta}=\frac{1}{2}(\hat{\varsigma}\rho_{\theta}+\rho_{\theta}\hat{\varsigma})$.
The QFI describes the maximal information on $\theta$ extractable from $\rho_{\theta}$ by optimizing all possible measurement observables. In our scheme, the quantity to be estimated is the gravitational acceleration $g$. The QFI of the MQ gravimeter can be evaluated from the density matrix $\rho_{1}=|\psi_{1}(t)\rangle\langle\psi_{1}(t)|$ in the decoherence-free case. Its explicit form is provided in the Supplementary Material~\cite{SM}. The QFI reaches its maximum at time $t_{1}=\pi/\omega$ as
\begin{equation}
\mathcal{F}^{g}_{1}=\frac{8m}{\hbar\omega^{3}}.\label{QFI1}
\end{equation}
It is found that the QFI increases with the particle mass, as visualized in Fig.~\ref{FIG2}(a) for different oscillator frequencies $\omega$. Repeating the experiments in a total time duration $\mathcal{T}$, we obtain the number of repeated experiments as $\upsilon=\mathcal{T}/t_1$. Then, the quantum Cram\'{e}r-Rao bound results in $\delta g_1=\omega\sqrt{{\hbar \pi/(8m\mathcal{T})}}$. Therefore, our MQ gravimetry can fully exploit the advantage enabled by the large mass of the suspended particle to improve the sensing performance. This is in sharp contrast to Refs.~\cite{PhysRevLett.111.180403,z8b4-sm79,2018QvarfortP36903690,2024LengP123601123601,2019GietkaP190801190801}, where the precision is independent of the mass. 

The QFI can be saturated by the Rabi measurement scheme. The obtained result gives the sensitivity of the scheme as $S_1=\sqrt{t}\delta O_1/|\partial \bar{O}_1/\partial g|$, where $\delta O_1=(\bar{O}_1-\bar{O}_1^2)^{1/2}$ is the standard deviation of the excited-state occupation operator $\hat{O}_1$ and $t$ is the sensing time~\cite{SM}. At the optimal sensing time $t=t_{1}$, the sensitivity achieves
\begin{equation}
S_{1}=\omega\sqrt{\frac{\hbar\pi}{8m}}.\label{S1}
\end{equation}
Exhibiting the same mass-dependent enhancement, Eq. \eqref{S1} reaches the one evaluated from the quantum Cram\'{e}r-Rao bound. It demonstrates that the used Rabi-measurement scheme is optimal for this system. The mass-dependent behavior of $S_{1}$ differs markedly from those in previous gravimetry schemes based on a levitated particle coupled to a qubit. For example, the sensitivities of a linear accumulation scheme and a recent gravity-sensing protocol, both employing a nitrogen-vacancy center as the auxiliary qubit, are given by $S_{\rm{S}}=\omega\sqrt{t_{\rm{S}}}/|2\gamma_{e}B't_{\rm{S}}|$~\cite{PhysRevLett.111.180403} and $S_{\rm{W}}=\omega\sqrt{t_{\rm{W}}}/|4\gamma_{e}B'\tau+2\gamma_{e}B'\omega T_{\rm{W}}|$~\cite{z8b4-sm79}, where $t_{\rm{S}}$ and $t_{\rm{W}}$ denote the respective sensing times.
Moreover, when the thermal noise is present, the dynamics of the system is governed by the master equation
\begin{equation}
\dot{\rho}_{1}(t)=-i/\hbar[\hat{\mathcal H}_{1}^{g},\rho_{1}(t)]+\kappa[(n_{\rm{th}}+1) \check{\mathcal{D}}_{\hat{\sigma}_{1}^{-}}+n_{\rm{th}} \check{\mathcal{D}}_{\hat{\sigma}_{1}^{+}}]\rho_{1}(t),   \label{mst} 
\end{equation}
where $\check{\mathcal{D}}_{\hat{o}}\cdot=\hat{o}\cdot\hat{o}^{\dagger}-\{\hat{o}^{\dagger}\hat{o},\cdot\}/2$ is the Lindblad superoperator, $\kappa=\omega/Q$ is the decay rate of the mechanical mode with the quality factor $Q$~\cite{SM}, and $n_{\rm{th}}=1/\{\exp[\hbar\omega/(k_{B}T)]-1\}$ is the thermal phonon number at temperature $T$, with $k_{B}$ the Boltzmann constant.
By numerically solving Eq. \eqref{mst}, we obtain the sensitivity $S_1^{\rm{diss}}$ at $t=t_1$ under the dissipative condition. Figure~\ref{FIG2}(b) compares the sensitivity $S_{1}$ and $S_1^{\rm{diss}}$ of our MQ gravimetry with $S_{\rm{S}}$ and $S_{\rm{W}}$. We find that both $S_{1}$ and $S_1^{\rm{diss}}$ surpass the mass-independent sensitivities $S_{\rm{S}}$ and $S_{\rm{W}}$ as the particle mass $m$ increases, thereby allowing higher-precision sensing of gravitational acceleration.

\section{The MCQ gravimeter}
\begin{figure}
\centering
\includegraphics[width=0.48\textwidth]{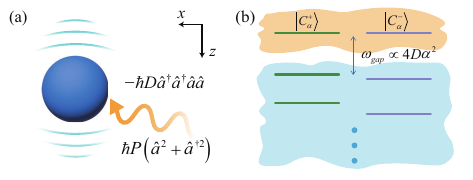}
\caption{(a) Schematic model of the MCQ gravimeter. (b) Energy spectrum of the MCQ.}\label{FIG3}
\end{figure}

To further improve the gravimetric precision beyond that achievable with the MQ gravimeter, we propose a MCQ gravimetry by combining the MQ architecture and mechanical cat states~\cite{PhysRevLett.130.213604}. It achieves a precision of double standard quantum limits. As shown in Fig.~\ref{FIG3}(a), we consider that a levitated particle in the presence of the Duffing nonlinearity is subject to a resonant two-phonon driving. In a frame rotating at frequency $\omega$, its free Hamiltonian reads $\hat{H}_{2}=-\hbar D\hat{a}^{\dagger}\hat{a}^{\dagger}\hat{a}\hat{a}+\hbar P(\hat{a}^{\dagger 2}+\hat{a}^{2})$, where $P$ denotes the driving strength. $\hat{H}_{2}$ has two degenerate ground states called the even and odd cat states $|C_{\alpha}^{\pm}\rangle=\mathbb{N}_{\alpha}^{\pm}(\left|+\alpha\right\rangle\pm \left|-\alpha\right\rangle)$, where $\left|\pm\alpha\right\rangle$ denote positive and negative coherent states with amplitude $\alpha=\sqrt{P/D}$ and $\mathbb{N}_{\alpha}^{\pm}=(2\pm 2e^{-2\alpha^2})^{-1/2}$ are normalization constants~\cite{2025DingP52795279,PhysRevX.9.041009,SM}. The Hilbert space is divided into a cat-state subspace spanned by $|C_{\alpha}^{\pm}\rangle$ and a residual subspace formed by the remaining eigenstates separated by an energy gap $\omega_{\rm{gap}}=4D\alpha^2$~\cite{SM}. Its energy spectrum is shown in Fig.~\ref{FIG3}(b). When external perturbations are smaller than the gap $\omega_{\rm{gap}}$, the dynamics of the system remains confined in the cat-state subspace. This defines the desired MCQ, with the Pauli operator $\hat{\sigma}_{2}^{x}=\left|C_{\alpha}^+\rangle\langle C_{\alpha}^-\right|+\left|C_{\alpha}^-\rangle\langle C_{\alpha}^+\right|$~\cite{PhysRevX.9.041009}. The presence of the gravity recasts the Hamiltonian of the levitated particle into
\begin{equation}
\hat{H}_{2}^{g}=\hat{H}_{2}+(mg-F)z_{0}(\hat{a}^{\dagger}e^{i\omega t}+\hat{a}e^{-i\omega t}).\label{H2g}
\end{equation}
Projected into the MCQ subspace, $\hat{H}_{2}^{g}$ becomes
\begin{equation}
\hat{\mathcal H}_{2}^{g} =\hbar \Omega_{2}\hat\sigma^{x}_{2}\cos{(\omega t)} \label{H2gC}
\end{equation}
with $\Omega_{2}\equiv 2\alpha (mg-F)z_{0}/\hbar$ as the effective Rabi frequency. Being the same as Fig.~\ref{FIG1}(d), our scheme based on the Rabi measurement protocol has the following steps. First, the MCQ is prepared in the state $|\psi_{2}(0)\rangle=|C_{\alpha}^-\rangle$, whose mean phonon number $N\simeq \alpha^2$ can be seen as the quantum-resource number. Then, governed by Eq. \eqref{H2gC}, the state evolves to $|\psi_{2}(t)\rangle=\cos{[\Omega_{2}/\omega\sin{(\omega t)}]}|C_{\alpha}^-\rangle-i\sin{[\Omega_{2}/\omega\sin{(\omega t)}]}|C_{\alpha}^+\rangle$. Third, a projective measurement to the even-cat-state occupation operator $\hat{O}_{2}=|C_{\alpha}^+\rangle\langle C_{\alpha}^+ |$ gives the mean value as $\bar{O}_{2}=\sin^2{[\Omega_{2}/\omega\sin{(\omega t)}]}$.

\begin{figure}
\centering
\includegraphics[width=0.48\textwidth]{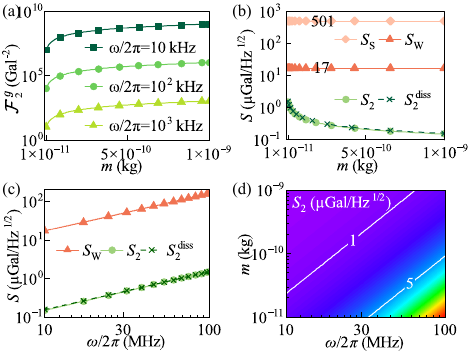}
\caption{(a) Mass-dependent variation of the QFI in the MCQ gravimetry scheme  with different frequency $\omega$. Sensitivities as a function of (b) the particle mass $m$ when $\omega/2\pi= 10~\text{kHz}$ and (c) the CM mode frequency $\omega$ when $m = 10^{-9}~\text{kg}$ for different sensing schemes. Parameters used to calculate $S_{2}^\mathrm{diss}$ are $Q=10^{8}$, $T=10~\rm{mK}$, and $\Omega_2=\pi\omega/4$. (d) Sensitivity $S_{2}$ for the MCQ gravimetry scheme as functions of $m$ and $\omega$. The shared parameters for all panels are $D=0.1\omega$ and $P=3.6\omega$. Logarithmic scales are used for the vertical axes in (a) and (b), and for both axes in (c) and (d). }\label{FIG4}
\end{figure}

The ultimate precision of the MCQ gravimeter quantified by the QFI is evaluated from the state $\rho_{2}=|\psi_{2}(t)\rangle\langle\psi_{2}(t)|$~\cite{SM}. The QFI reaches its maximum at sensing time $t_{2}=\pi/(2\omega)$ as
\begin{equation}
\mathcal{F}^{g}_{2}=\frac{8Nm}{\hbar\omega^{3}}.\label{QFI2}
\end{equation}
It reveals that the QFI linearly increases with not only the particle mass $m$ as shown in Fig.~\ref{FIG4}(a), but also the mean phonon number $N$. The quantum Cram\'{e}r-Rao bound gives the root-mean-square error as $\delta g_2=\omega\sqrt{{\hbar \pi/(16Nm\mathcal{T})}}$. Both the scaling rules of $\delta g_2$ with $N$ and $m$ reach the standard quantum limit~\cite{RevModPhys.89.035002,2021WuP5404254042,2019LiuP2300123001}, which we refer to as the double standard quantum limits. Thus, the MCQ gravimeter characterized by Eq. \eqref{QFI2} outperforms the MQ one characterized by Eq. \eqref{QFI1} in an enhanced ratio $N$. This is further confirmed by the sensitivity of the MCQ sensing scheme at the sensing time $t_{2}$ explicitly obtained under our Rabi-measurement scheme~\cite{SM},
\begin{equation}
S_{2}=\omega\sqrt{\frac{\hbar\pi}{16Nm}},\label{S2}
\end{equation}
which is derived from the MCQ flip probability $\bar{O}_{2}$. Equation \eqref{S2} reveals that $S_{2}$ saturates the QFI $\mathcal{F}^{g}_{2}$ from Eq.~\eqref{QFI2}, indicating that the sensing scheme for the MCQ gravimetry is optimal.

Beyond the ideal case, we also analyze the effect of  thermal noise on the sensitivity of the MCQ gravimetry. The density matrix $\rho_{2}(t)$ of the system obeys the master equation in the large-$N$ limit
\begin{equation}
\dot{\rho}_{2}(t)=-i/\hbar[\hat{\mathcal H}_{2}^{g},\rho_{2}(t)]+\kappa(2n_{\rm{th}}+1) {N}\check{\mathcal{D}}_{\hat{\sigma}_{2}^{x}}\rho_{2}(t). \label{ME}
\end{equation}
Solving Eq. \eqref{ME} under the initial condition $|\psi_2(0)\rangle$, the mean value of the even-cat-state occupation operator at the sensing time $t_{2}$ is obtained as $\bar{O}_{2}=1/2-1/2\cos{[2\Omega_{2}/\omega\sin{(\omega t_{2})}]}\exp{[-2\kappa(2n_{\rm{th}}+1) N t_{2}]}$, from which the corresponding sensitivity~\cite{SM} at $t_{2}$  and $\Omega_2=\pi\omega/4$  is obtained as
\begin{equation}
S_{2}^\text{diss}=\omega\sqrt{\frac{\hbar\pi}{16Nm}}e^{2\kappa(2n_{\rm{th}}+1)Nt_{2}}. \label{S2diss}
\end{equation}
Figure~\ref{FIG4}(b) shows the sensitivities $S_{2}$ and $S_{2}^\text{diss}$ of the MCQ gravimetry scheme as functions of particle mass $m$ under ideal conditions and in the presence of thermal noise with the temperature $T=10~\rm{mK}$, respectively, alongside the sensitivities $S_{\rm{S}}$ and $S_{\rm{W}}$ of previous approaches.
It is evident from Fig.~\ref{FIG4}(b) that increasing the particle mass $m$ allows the MCQ gravimeter to reach a sensitivity about $0.1~\text{\textmu}\text{Gal}/\sqrt{\text{Hz}}$ even in the presence of thermal noise, outperforming the single-particle sensitivities reported in traditional schemes with a levitated particle~\cite{PhysRevLett.111.180403,z8b4-sm79,2024LengP123601123601} by two orders of magnitude. Furthermore, although the sensitivities $S_{2}$ and $S_{2}^\text{diss}$ of the MCQ gravimetry scheme increase with increasing oscillator frequency $\omega$, indicating a slight deterioration in performance, they remain lower than the sensitivities $S_{\rm{W}}$ reported in recent scheme across a wide frequency range, as presented in Fig.~\ref{FIG4}(c). Figure~\ref{FIG4}(d) illustrates the variation of the sensitivity $S_{2}$ of the MCQ gravimeter under the combined influence of particle mass $m$ and oscillator frequency $\omega$, showing that $S_{2}$ can be significantly reduced within experimentally achievable parameter ranges. Additionally,  a comparison of sensitivities and parameters for different schemes are provided in the Supplementary Material~\cite{SM}. These results show that the MCQ gravimeter not only fully exploits the advantage of increased particle mass, but also approaches the double standard quantum limits.

\section{Experimental feasibility} MQs can be constructed by introducing Duffing nonlinearity into mechanical oscillators, as has been experimentally demonstrated~\cite{2024YangP783788}. A similar nonlinearity can be induced in the CM mode of levitated particles in our proposed gravimetry scheme by engineering a trapping potential of the form $z^4$, as demonstrated in Ref.~\cite{2022ZhouP4650}. In addition, cat states for mechanical modes have been successfully prepared, which can be mapped to the CM mode of a particle with mass $3\times10^{-11}~\text{kg}$ ~\cite{doi:10.1126/science.adf7553}. Two-phonon driving of the center-of-mass modes is achieved via the direct application of a cosine-waveform, time-varying electric field for electrically levitated particles~\cite{doi:10.1126/science.aaw2884,PRXQuantum.5.020314,PhysRevLett.130.073602,PhysRevLett.122.030501}. For magnetically levitated systems, it entails positioning a wire near the magnet to carry a similarly modulated current~\cite{PhysRevApplied.8.034002,2019TimberlakeP224101224101,PhysRevLett.130.073602,PhysRevA.107.023722}.
Moreover, the levitated particles are highly isolated from the environment, allowing their CM modes to achieve a quality factor of $10^{8}$ ~\cite{PhysRevLett.132.133602,PhysRevLett.100.031101,PhysRevLett.109.103603,2013GieselerP806810,PhysRevLett.124.093602}, which can be achievable at low temperature $T=10~\rm{mK}$ \cite{2019KosenP101120101120,doi:10.1126/science.abc7312} in our scheme~\cite{SM}. We consider a mechanical mode with a vibration frequency of $\omega/2\pi=10~\text{kHz}$ and $m=10^{-9}~\text{kg}$, featuring a Duffing nonlinearity of strength $D/2\pi=1~\text{kHz}$, under a static force $F\sim9.8\times10^{-9}~\rm{N}$. The two-phonon driving strength is chosen as $P=3.6\omega$. In comparison with previous schemes shown in Figs.~\ref{FIG3}(b), \ref{FIG4}(b), and \ref{FIG4}(c), the parameters used to compute $S_{\rm{S}}$ and $S_{\rm{W}}$ and the resulting values are consistent with the data in Table I of Ref.~\cite{z8b4-sm79}. Under the parameters considered here, our approaches reduce the gravitational sensitivity by two orders of magnitude compared with previous methods~\cite{z8b4-sm79}, with further improvement achievable under experimentally feasible conditions.

\section{Conclusion} To overcome the intrinsic limitations of conventional quantum gravimeters, we propose a quantum gravimeter based on MQs and MCQs of a levitated mesoscopic particle, which obviates the need for additional quantum sensors and directly harnesses the large particle mass, enabling gravimetry at the double standard quantum limits. Theoretical analysis indicates that this approach can achieve a gravitational sensitivity on the order of $0.1~\text{\textmu}\text{Gal}/\sqrt{\text{Hz}}$, outperforming the traditional schemes by two orders of magnitude. This is of broad relevance to both fundamental research and engineering applications, spanning the exploration of quantum physics at the mesoscopic scale~\cite{doi:10.1126/sciadv.adk2949},  the  investigation of geological hazards~\cite{https://doi.org/10.1029/2022GL097814,doi:10.1126/science.1128661,2016MontagnerP1334913349}, as well as gravity cartography~\cite{2022StrayP590594,2026LiP6262} and geophysical surveying \cite{2026JuniorP,2025MigliaccioP865925,2021ForsterP2626,2018PearsonGrantP146157}.

\begin{acknowledgments}
P.B.L. is supported by the National Natural Science Foundation of China under Grants No. W2411002 and No. 12375018. J.H.A. is supported by the National Natural Science Foundation of China (Grants No. 12275109, No. 92576202, and No. 12247101), the Quantum Science and Technology-National Science and Technology Major Project (Grant No. 2023ZD0300904), and the Gansu Science and Technology Leading Talent Program (Grant No. 26RCKA011).
\end{acknowledgments}

\bibliography{ArticleReference}
\end{document}


\title{Supplemental Materials for ``Quantum gravimetry with mechanical qubits''}
\author{Xiao-Wen Huo}
\affiliation{Ministry of Education Key Laboratory for Nonequilibrium Synthesis and Modulation of Condensed Matter, Shaanxi Province Key Laboratory of Quantum Information and Quantum Optoelectronic Devices, School of Physics, Xi'an Jiaotong University, Xi'an 710049, China}
\author{Jun-Hong An}
\affiliation{Key Laboratory of Quantum Theory and Applications of MoE, Lanzhou Center for Theoretical Physics, and Key Laboratory of Theoretical Physics of Gansu Province, Lanzhou University, Lanzhou 730000, China}
\author{Peng-Bo Li}
\email{lipengbo@mail.xjtu.edu.cn}
\affiliation{Ministry of Education Key Laboratory for Nonequilibrium Synthesis and Modulation of Condensed Matter, Shaanxi Province Key Laboratory of Quantum Information and Quantum Optoelectronic Devices, School of Physics, Xi'an Jiaotong University, Xi'an 710049, China}

\date{\today}

\begin{abstract}
In this supplemental material, we first provide more details about the mechanical qubit (MQ) gravimeter in Sec.~\ref{sec0},  including its Hamiltonian, quantum Fisher information (QFI), and sensitivity.
Next, we show how to realize the mechanical cat qubit (MCQ) and its quantum properties, and then we also present the Hamiltonian, QFI and sensitivity of the MCQ gravimeter, in Sec.~\ref{secI} and Sec.~\ref{secII}, respectively.
Additionally, for the MCQ gravimetry scheme, we examine the sensitivity under single-phonon loss and white-noise conditions, in addition to the ideal case.
Finally, we analyze the dominant sources of dissipation in the the center-of-mass (CM) vibrational mode of the levitated mesoscopic particle, in Sec.~\ref{secIII}. 
\end{abstract}
\maketitle

\tableofcontents
\newpage

\section{\label{sec0}Mechanical-qubit gravimeter}
This section demonstrates details of the mechanical qubit (MQ) gravimeter. The Hamiltonians of the MQ and the corresponding gravimeter are first given. The QFI, the sensitivity, and the optimal sensing time of the MQ gravimetry scheme are then derived. 
\subsection{Hamiltonian}
The Hamiltonian of a levitated mesoscopic particle is  $\hat{H}_{1}=\hbar\omega\hat{a}^{\dagger}\hat{a}-\hbar D\hat{a}^{\dagger}\hat{a}^{\dagger}\hat{a}\hat{a}$, where $\hat{a}$ is the annihilation operator of the center-of-mass (CM) mode of the particle with frequency $\omega$ along the direction of gravity, and $D$ characterizes the strength of the Duffing nonlinearity. The energy differences between the second excited state $|2\rangle$ and the first excited state $|1\rangle$, and between $|1\rangle$ and the ground state $|0\rangle$, of the system are given by $E_{21}=\hbar\omega-2\hbar D$ and $E_{10}=\hbar\omega$, respectively. The anharmonicity of this system defined as $\Delta_{1}=|E_{21}-E_{10}|$ is therefore $2\hbar D$.
Provided that external perturbations are much smaller than the anharmonicity $\Delta_{1}$, the system is restricted to the subspace spanned by $|0\rangle$ and $|1\rangle$. This forms a MQ, with $|0\rangle$ and $|1\rangle$ being the ground and excited states of the MQ. Based on this definition, the Pauli operators of the MQ reads $\hat{\sigma}_{1}^{x}=|1\rangle\langle0|+|0\rangle\langle1|$, $\hat{\sigma}_{1}^{y}=-i|1\rangle\langle0|+i|0\rangle\langle1|$, and $\hat{\sigma}_{1}^{z}=|1\rangle\langle1|-|0\rangle\langle0|$ and the projection operator is defined as $\hat{\mathcal{P}}_{1}=|1\rangle\langle1|+|0\rangle\langle0|$.

Under the influence of gravity, the system is described by the Hamiltonian
\textbf{\begin{equation}
				\hat{H}^{g}_{1}=\hbar\omega\hat{a}^{\dagger}\hat{a}-\hbar D\hat{a}^{\dagger}\hat{a}^{\dagger}\hat{a}\hat{a}+(mg-F)z_{0}(\hat{a}^{\dagger}+\hat{a}),
\end{equation}}
$z_{0}=\sqrt{\hbar/(2m\omega )}$ is the zero-point fluctuation. $F$ is a static force along the direction opposite to gravity to ensure $(mg-F)z_{0}\ll\Delta_{1}$, so that the system remains confined in the computational subspace. To project $\hat{H}^{g}_{1}$ onto the MQ subspace, the projection operator $\hat{\mathcal{P}}_1$ is applied.
Since the projection operator satisfies
\begin{equation}
\hat{\mathcal{P}}^{\dagger}_{1}\hat{a}^{\dagger}\hat{\mathcal{P}}_{1}=|1\rangle\langle0|=\sigma_{1}^{+},~\hat{\mathcal{P}}^{\dagger}_{1}\hat{a}\hat{\mathcal{P}}_{1}=|0\rangle\langle1|=\sigma_{1}^{-}, 
\end{equation}
the projected Hamiltonian is obtained as
\begin{equation}
\hat{\mathcal{H}}^{g}_{1}=\hat{\mathcal{P}}^{\dagger}_{1}\hat{H}^{g}_{1}\hat{\mathcal{P}}_{1}=\hbar\frac{\omega}{2}\hat{\sigma}_{1}^{z}+\hbar\Omega_{1}\hat{\sigma}_{1}^{x}
\end{equation}
with $\Omega_{1}\equiv(mg-F)z_{0}/\hbar$.
This Hamiltonian describes the MQ gravimeter employed in the subsequent gravity-sensing protocol.

\subsection{Quantum Fisher information and Sensitivity}
The quantum Fisher information (QFI) sets the lower bound on the variance of a measured parameter in quantum precision measurements, representing the optimal performance achievable by the sensing system. For a quantum sensor described by the density matrix $\rho_{\theta}$, where $\theta$ is the parameter to be estimated, the QFI is defined as $\mathcal{F}^{\theta}=\mathrm{Tr}[\hat{\varsigma}^{2}\rho_{\theta}]$, where the symmetric logarithmic derivative $\hat{\varsigma}$ satisfies $\partial_{\theta}\rho_{\theta}=(\hat{\varsigma}\rho_{\theta}+\rho_{\theta}\hat{\varsigma})/2$~\cite{2021WuP5404254042,2019LiuP2300123001}. In the diagonal representation of the density matrix, $\rho_{\theta}^{\Lambda}=\sum_{j}\lambda_{j}|\lambda_{j}\rangle$, the QFI is expressed $\mathcal{F}^{\theta}=\text{Tr}[\hat{\varsigma}^{2}\rho_{\theta}^{\Lambda}]$, with $\hat{\varsigma}=\sum_{j,k:\lambda_{j}+\lambda_{k}\ne0}\frac{2(\partial_{\theta}\rho_{\theta})^{\Lambda}_{jk}|\lambda_{j}\rangle\langle\lambda_{k}|}{\lambda_j+\lambda_k}$. 
For the MQ gravimeter, the ideal dynamical equation $\dot{\rho}_{1}=-i/\hbar[\hat{\mathcal{H}}^{g}_{1},\rho_{1}]$ is solved under the initial state $\rho_{1}=|0\rangle\langle0|$ as 
\begin{equation}
	\rho_{1}(t)= \begin{bmatrix}
		 \frac{2\Omega_{1}^{2}}{\mathcal{A}^{2}}[1-\cos(\mathcal{A}t)]&\frac{\Omega_{1}}{\mathcal{A}^{2}}[-\omega+\omega\cos(\mathcal{A}t)-i\mathcal{A}\sin(\mathcal{A}t)] \\
		\frac{\Omega_{1}}{\mathcal{A}^{2}}[-\omega+\omega\cos(\mathcal{A}t)+i\mathcal{A}\sin(\mathcal{A}t)]&\frac{\omega^{2}+2\Omega_{1}^{2}+2\Omega_{1}^{2}\cos(\mathcal{A}t)}{\mathcal{A}^{2}}
	\end{bmatrix},
\end{equation}
with $\mathcal{A}=\sqrt{\omega^{2}+4\Omega_{1}^{2}}$.
Diagonalization of $\rho_{1}$ yields $\rho_{1}^{\Lambda}(t)=\begin{bmatrix}
	1&0 \\
	0&0
\end{bmatrix}$ in the eigenbasis ${|\lambda_{1/2}\rangle}$ with eigenvalues $\lambda_1=1$ and $\lambda_2=0$.
The transformation matrix that diagonalizes $\rho_1(t)$ is given by
\begin{equation}
	\Lambda_{1}= \begin{bmatrix}
		\frac{2\Omega_{1}}{\mathcal{A}}\sin(\mathcal{A}t/2)&\frac{\omega}{\mathcal{A}}\sin(\mathcal{A}t/2)+i\cos(\mathcal{A}t/2) \\
		-\frac{\omega}{\mathcal{A}}\sin(\mathcal{A}t/2)+i\cos(\mathcal{A}t/2)&\frac{2\Omega_{1}}{\mathcal{A}}\sin(\mathcal{A}t/2)
	\end{bmatrix}.
\end{equation}
Next, we compute the symmetric logarithmic derivative operator $\hat{\varsigma}$.
The derivative of the density matrix with respect to gravitational acceleration $g$ in the diagonal basis  is obtained as
\begin{equation}
	(\partial_{g}\rho_{1})^{\Lambda}=\Lambda_{1}^{\dagger}(\partial_{g}\rho_{1})\Lambda_{1}=
	\begin{bmatrix}
		0&l_{12} \\
		l_{21}&0
	\end{bmatrix},
\end{equation}
where $l_{12}$ and $l_{21}$ denote the corresponding nonzero matrix elements.
It leads to 
\begin{equation}
	\hat{\varsigma}=
	2\begin{bmatrix}
		0&l_{12} \\
		l_{21}&0
	\end{bmatrix}.
\end{equation}
Finally, substituting $\hat{\varsigma}$ and $\rho_1^{\Lambda}(t)$ into the definition of the QFI yields
\begin{equation}
	\mathcal{F}_{1}=\frac{8m^2z_{0}}{\hbar^{2}\mathcal{A}^{6}}\{32\Omega_{1}^{2}t^2+\Omega_{1}^{2}\omega^2(2+8\Omega_{1}^{2}t^{2})+\omega^{4}-\omega^2(4\Omega_{1}^{2}+\omega^{2})\cos(\mathcal{A}t)+\Omega_{1}^{2}\omega^2[\cos(2\mathcal{A}t)+4\mathcal{A}t\sin(\mathcal{A}t)]\},
\end{equation}
quantifies the maximum information about the gravitational acceleration $g$ obtainable from the MQ gravimeter.

\begin{figure}
    \centering
    \includegraphics[width=0.6\textwidth]{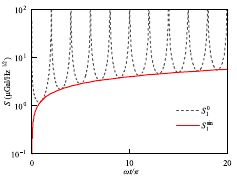}
    \caption{Temporal variation of the sensitivity $S_{1}^{0}$ of the MQ gravimeter (black curve) and its envelope (red curve). The parameters are chosen as $\omega/2\pi=10~\rm{kHz}$, $D=0.1\omega$, $m=10^{-9}~\rm{kg}$, and $\Omega_{1}=0.02\omega$.}
    \label{SMFIG1}
\end{figure}

In addition to characterizing the performance through the QFI, the MQ gravimetry scheme is further evaluated in terms of sensitivity.
Our gravity sensing protocol is based on the Rabi measurement scheme, in which the signal to be estimated is encoded in the flip probability of the MQ.
The sensitivity $S$ is therefore derived from the flip probability.
It is defined as $S=\sqrt{t}\delta O/|\partial \bar{O}/\partial g|$, where $\delta O=\sqrt{\bar{O}-\bar O^{2}}$ denotes the standard deviation of the state
occupation operator $\hat{O}$, and $t$ is the sensing time.
For the MQ gravimetry scheme,  $\hat{O}_1=|1\rangle\langle1|$ is the excited-state
occupation operator, and the initial state is taken as $|\psi_{1}(0)\rangle=|0\rangle$.
Under ideal conditions, the time evolution operator $\hat{U}_{1}(t)=e^{-i\hat{\mathcal H}^{g}_{1}t/\hbar}$ generates the state $|\psi_{1}(t)\rangle=\left[\cos{(\mathcal{A}t/2)}+i\omega/\mathcal{A}\sin{(\mathcal{A}t/2)}\right]|0\rangle-2i\Omega_{1}/\mathcal{A}\sin{(\mathcal{A}t/2)}|1\rangle$.
The corresponding  mean value of the  excited-state
occupation operator $\hat{O}_1$ is given by $\bar{O}_1=4\Omega_{1}^{2}/\mathcal{A}^{2}\sin^{2}{(\mathcal{A}t/2)}$.
The standard deviation of $\hat{O}_1$ and its derivative with respect to $g$ are obtained as
\begin{equation}
	\begin{split}
		\delta O_{1}&=\frac{2\Omega_{1}}{\mathcal{A}^{2}}\sin(\mathcal{A}t/2)\sqrt{\omega^2+2\Omega_{1}^{2}+\Omega_{1}^{2}\cos(\mathcal{A}t)}, \\
		\frac{\partial \bar{O}_{1}}{\partial g}&=\frac{8mz_{0}\Omega_{1}}{\hbar\mathcal{A}^{4}}\sin(\mathcal{A}t/2)[2\Omega_{1}^{2}\mathcal{A}t\cos(\mathcal{A}t/2)+\omega^2\sin(\mathcal{A}t/2)].
	\end{split}
	\label{p1}
\end{equation}
Substituting Eq.~\eqref{p1} into the definition of the sensitivity yields
\begin{equation}
		S_{1}^{0}=\sqrt{t}\mathcal{A}^{2}\sqrt{\frac{\hbar\omega}{8m}}\frac{\sqrt{\omega^{2}+4\Omega_{1}^{2}\cos^{2}(\mathcal{A}t/2)}}{|2\Omega_{1}^{2}\mathcal{A}t\cos(\mathcal{A}t/2)+\omega^{2}\sin(\mathcal{A}t/2)|}.
	\label{S10}
\end{equation}
Figure~\ref{SMFIG1} shows the sensitivity $S_{1}^{0}$ as a function of sensing time $t$ by the black dashed line.
It is observed that $S_{1}^{0}$ exhibits periodic oscillations in time.
The envelope of this oscillatory behavior is therefore used to determine the optimal sensing time $t_{1}$.
When the condition $\mathcal{A}t=(2n+1)\pi$, with $n\in\mathbb{N}$, is satisfied, the sensitivity reduces to $S_{1}^{\rm{sin}}=\sqrt{t}\mathcal{A}^{2}/\omega\sqrt{\hbar\omega/(8m)}$. The red curve in Fig.~\ref{SMFIG1} shows $S_{1}^{\rm{sin}}$. It exactly corresponds to the envelope of the sensitivity $S_{1}^{0}$ given by Eq.~\eqref{S10}.
Therefore, the optimal sensing time is obtained for $n=0$, i.e., when $\mathcal{A}t=\pi$.

Under the condition $\Omega_{1}\ll\Delta_{1}$, which ensures that the MQ operates within the computational subspace, one has $\mathcal{A}\approx\omega$.
The optimal sensing time is therefore $t_{1}=\pi/\omega$ at which the sensitivity becomes
\begin{equation}
	S_{1}=\omega\sqrt{\frac{\hbar\pi}{8m}}.
	\label{S1}
\end{equation}
Additionally, the QFI evaluated at the optimal sensing time $t_{1}=\pi/\omega$ reduces to
\begin{equation}
	\mathcal{F}_{1}^{g}=\frac{8m}{\hbar\omega^3}.
	\label{F1G}
\end{equation}
It is found that $S_{1}$ is consistent with $\mathcal{F}_{1}^{g}$, indicating that our sensing scheme is the optimal sensing scheme.
Importantly,  both the sensitivity $S_{1}$  and QFI $\mathcal{F}_{1}^{g}$  depend on particle mass $m$, with performance improving as particle mass increases. This feature highlights a key advantage of the levitated-particle MQ gravimetry scheme.

\section{\label{secI}Parametrically driven Duffing mechanical oscillator}
In this section, we analyze the quantum properties of the parametrically driven duffing mechanical oscillator.
We first present its energy spectrum and construct a mechenical cat qubit (MCQ).  
Subsequently, we demonstrate that the MCQ offers significant advantages over the MQ in suppressing leakage from the qubit subspace and mitigating phase-flip errors.

\subsection{Energy spectrum and MCQ}
\begin{figure}
	\centering
	\includegraphics[width=0.6\textwidth]{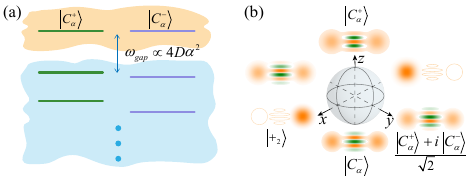}
	\caption{(a) The eignenergy spectrum of the MCQ. (b) Wigner function distributions of the MCQ and their representations on the Bloch sphere.}
	\label{SMFIG2}
\end{figure}

In a frame rotating at the mechanical mode frequency $\omega$, the Hamiltonian of a parametrically driven duffing mechanical oscillator is 
\begin{equation}
\hat{H}_{2}=-\hbar D\hat{a}^{\dagger}\hat{a}^{\dagger}\hat{a}\hat{a}+\hbar P(\hat{a}^{\dagger 2}+\hat{a}^{2}),
\label{H2}
\end{equation}
where $P$ denotes the strength of the resonant two-phonon drive.
$\hat{H}_{2}$ commutes with the phonon parity operator $\exp(i\pi\hat{a}^{\dagger}\hat{a})$, i.e., $[\hat{H}_{2},\exp(i\pi\hat{a}^{\dagger}\hat{a})]=0$. Thus, its eigenspace is divided into the even- and odd-parity manifolds, indicated by green and purple lines in Fig.~\ref{SMFIG2}(a), respectively.

Below, we derive the eigenstates of $\hat{H}_{2}$.
The action of displacement operators $\hat{\mathcal{D}}(\pm\alpha)=\exp(\pm\alpha\hat{a}^{\dagger}\mp\alpha\hat{a})$, with $\alpha=\sqrt{P/D}$, converts $\hat{H}_{2}$ into
\begin{equation}
	\hat{H}_{2}^{\prime}=\hat{\mathcal{D}}(\pm\alpha)\hat{H}_{2}\hat{\mathcal{D}}^{\dagger}(\pm\alpha)=-4\hbar D\alpha^2\hat{a}^{\dagger}\hat{a}-\hbar D\hat{a}^{\dagger}\hat{a}^{\dagger}\hat{a}\hat{a}\mp2\hbar D\alpha(\hat{a}^{\dagger}\hat{a}\hat{a}+\hat{a}^{\dagger}\hat{a}^{\dagger}\hat{a}),
	\label{H21}
\end{equation}
with constant energy shifts omitted.
The vacuum state $|0\rangle$ is clearly an eigenstate of $\hat{H}_{2}^{\prime}$. 
Therefore, the coherent state $|\mp\alpha\rangle=\hat{\mathcal{D}}(\mp\alpha)|0\rangle$ are eigenstates of $\hat{H}_{2}$.
After making linear combinations of $|\mp\alpha\rangle$, we obtain the even-odd cat states
$|C_{\alpha}^{\pm}\rangle=\mathbb{N}_{\alpha}^{\pm}(\left|+\alpha\right\rangle\pm \left|-\alpha\right\rangle)$, with $\mathbb{N}_{\alpha}^{\pm}=1/\sqrt{2[1\pm\exp(-2|\alpha|^{2})]}$, which are orthogonal degenerate eigenstates with the same eigenvalue of $\hat{H}_{2}$.
In the large-$\alpha$ limit, Eq. \eqref{H21} reduces to $\hat{H}_{2}^{\prime}\simeq-4\hbar D\alpha^2\hat{a}^{\dagger}\hat{a}$, whose eigenstates are the Fork states $|n\rangle$. 
After the inverse displacement transformation, we have the eigenstates of $\hat{H}_2$ as $|\phi_{\alpha,n}^{\mp}\rangle=N_{\alpha,n}^{\mp}[\hat{\mathcal{D}}(\alpha)\mp\hat{\mathcal{D}}(-\alpha)]|n\rangle$ in the large-$\alpha$ limit.
The ground states with $n=0$ just give $|\phi_{\alpha,0}^{\pm}\rangle=|C_{\alpha}^{\pm}\rangle$, which forms the basis of the two-dimensional cat-state subspace. The energy gap between the cat-state subspace and the remainder subspaces scales as  $\omega_{\rm{gap}}\propto4D\alpha^2$~\cite{2019PuriP4100941009}. 
When external perturbations are far smaller than the energy gap $\omega_{\rm{gap}}$, the system remains confined within the cat-state subspace depicted as the orange blocks in Fig.~\ref{SMFIG2}(a). Thus, the system becomes a MCQ. Its Pauli operators are $\hat{\sigma}_{2}^{x}=|C_{\alpha}^{+}\rangle\langle C_{\alpha}^{-}|+|C_{\alpha}^{-}\rangle\langle C_{\alpha}^{+}|$, $\hat{\sigma}_{2}^{y}=-i|C_{\alpha}^{+}\rangle\langle C_{\alpha}^{-}|+i|C_{\alpha}^{-}\rangle\langle C_{\alpha}^{+}|$, and $\hat{\sigma}_{2}^{z}=|C_{\alpha}^{+}\rangle\langle C_{\alpha}^{+}|-|C_{\alpha}^{-}\rangle\langle C_{\alpha}^{-}|$. The projection operator to the cat-state subspace is defined as $\hat{\mathcal{P}}_{2}=|C_{\alpha}^{+}\rangle\langle C_{\alpha}^{+}|+|C_{\alpha}^{-}\rangle\langle C_{\alpha}^{-}|$.
The Wigner functions and the corresponding Bloch-sphere distributions of the eigenstates of these operators are shown in Fig.~\ref{SMFIG2}(b).
The even and odd cat states $|C_{\alpha}^{\pm}\rangle$, which constitute the $Z$-eigenstates of the encoded MCQ, are located at the north and south poles of the Bloch sphere.  
Accordingly, the phase-space Wigner function representations of the eigenstates of the three Pauli operators ($\sigma_{2}^{x}$, $\sigma_{2}^{y}$ and $\sigma_{2}^{z}$) are shown near their corresponding positions on the Bloch sphere, as illustrated in Fig.~\ref{SMFIG2}(b).

\subsection{Superiority of the mechanical cat qubit}
The MCQ provides significant advantages over conventional MQ in suppressing the leakage from the qubit subspace and mitigating phase-flip errors.

\begin{figure}
	\centering
	\includegraphics[width=0.6\textwidth]{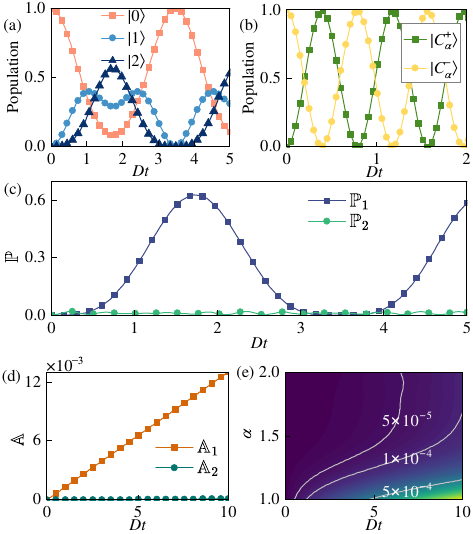}
	\caption{(a)	and (b) show the population dynamics of the MQ and the MCQ, respectively.
		(c)  Dynamics of the leakage probabilities $\mathbb{P}_{1}$ for the MQ (blue line) and $\mathbb{P}_{2}$ for the MCQ (green line).
		(d) Dynamics of phase-flip errors $\mathbb{A}_{1}$ for  the MQ (orange line) and $\mathbb{A}_{2}$ for the MCQ (green line). The parameters used above is $\alpha=2$.
		(e) $\mathbb{A}_{2}$  under the combined influence of evolution time $t$ and parameter $\alpha$.
		The parameters used to calculate $\mathbb{A}$ are $Q=10^8$, $\omega/2\pi=10~\rm{kHz}$, and $D=0.1\omega$.}
	\label{SMFIG3}
\end{figure}

To characterize the performance of the MCQ, we calculate the anharmonicity defined as the difference between the two lowest adjacent transition frequencies.
The magnitude of anharmonicity determines the probability of leakage from the qubit subspace under external perturbations. A larger anharmonicity suppresses the leakage, thereby making the proposed qubit more robust.
Based on the above discussion, we can clearly see that the anharmonicity of MCQ is $\Delta_2=4D\alpha^2$, which is significantly greater $\Delta_1=2D$ of the conventional MQ.
Therefore, the MCQ exhibits markedly improved performance over the MQ in suppressing leakage from the qubit subspace.
To verify this, we apply a resonant drive $\hbar\omega_{d} (\hat{a}^\dagger+
\hat{a})$ with equal amplitudes $\omega_{d}$ on both the MQ and the MCQ. 
In the rotating frame at the mechanical oscillator frequency $ω$, their respective Hamiltonians are
\begin{equation}
	\begin{split}
		\hat{H}_{1}^{d}&=-\hbar D\hat{a}^{\dagger}\hat{a}^{\dagger}\hat{a}\hat{a}+\hbar\omega_{d}(\hat{a}^{\dagger}+\hat{a}), \\
		\hat{H}_{2}^{d}&=-\hbar D\hat{a}^{\dagger}\hat{a}^{\dagger}\hat{a}\hat{a}+\hbar P(\hat{a}^{\dagger 2}+\hat{a}^{2})+\hbar\omega_{d}(\hat{a}^{\dagger}+\hat{a}).
	\end{split}
\end{equation}
We plot the time evolution of energy-level populations for both qubits under this driving in Fig.~\ref{SMFIG3}(a) and \ref{SMFIG3}(b), with $\omega_{d}=D$. Figure. ~\ref{SMFIG3}(a) shows that the MQ occupies not only the ground state $|0\rangle$ and the first excited state $|1\rangle$, but also the second excited state $|2\rangle$. It indicates an evident leakage from the qubit subspace. In contrast, Fig.~\ref{SMFIG3}(b) shows that as the population of the odd cat state $|C_{\alpha}^-\rangle$ decreases from unity to zero, the one of the even cat state $|C_{\alpha}^+\rangle$ increases from zero to unity. This reveals that no population in the residual subspaces is excited and thus no leakage occurs for the MCQ.
Fig.~\ref{SMFIG3}(c) presents the time evolution of the leakage probabilities $\mathbb{P}_{1}=1-\sum_{l=0,1}\langle l|\tilde{\rho}_{1}|l\rangle$ and $\mathbb{P}_{2}=1-\sum_{o=\pm}\langle C_{\alpha}^{o}|\tilde{\rho}_{2}|C_{\alpha}^{o}\rangle$ for both qubits, where $\tilde{\rho}_{1/2}$ are the density matrices of the MQ and MCQ, respectively.
The comparison clearly indicates that the MCQ exhibits an excellent suppression of leakage from the cat-state subspace.

Furthermore, the MCQ exhibits a pronounced advantage over the MQ in suppressing phase-flip errors.
The dynamics of the MQ and our MCQ, including single-phonon loss to the environment, are governed by the master equations $\dot{\tilde{\rho}}_{1}^{\rm{diss}}=-i/\hbar[\hat{H}_{1},\tilde{\rho}_{1}^{\rm{diss}}]+\kappa\check{\mathcal{D}}_{\hat{a}}\tilde{\rho}_{1}^{\rm{diss}}$ and $\dot{\tilde{\rho}}_{2}^{\rm{diss}}=-i/\hbar[\hat{H}_{2},\tilde{\rho}_{2}^{\rm{diss}}]+\kappa\check{\mathcal{D}}_{\hat{a}}\tilde{\rho}_{2}^{\rm{diss}}$, where $\kappa=\omega/Q$ is the decay rate of the mechanical mode with the quality factor $Q$, and $\check{\mathcal{D}}_{\hat{o}}\cdot=\hat{o}\cdot\hat{o}^\dagger-\{\hat{o}^\dagger\hat{o},\cdot\}/2$.
Assuming that the coupling between the mechanical mode and the environment remains sufficiently weak, the systems' dynamics are effectively restricted  to the computational subspaces.
First, applying the projection operator $\hat{\mathcal{P}}_{2}$ to the annihilation and creation operators yields
\begin{equation}
\begin{split}
\hat{\mathcal{P}}_{2}^{\dagger}\hat{a}^{\dagger}\hat{\mathcal{P}}_{2}=\alpha\sigma_{2}^{x}-i\alpha e^{-2|\alpha|^{2}}\sigma_{2}^{y}, ~
\hat{\mathcal{P}}_{2}^{\dagger}\hat{a}\hat{\mathcal{P}}_{2}=\alpha\sigma_{2}^{x}+i\alpha e^{-2|\alpha|^{2}}\sigma_{2}^{y}.
\end{split}\label{P2}
\end{equation}
in the large-$\alpha$ limit.
In this regime, the master equation is projected onto the cat-state subspace, yielding
\begin{equation}
\dot{\tilde{\rho}}_{2}^{\rm{diss}}=-i/\hbar[\hat{H}_{2},\tilde{\rho}_{2}^{\rm{diss}}]+\kappa\alpha^{2}\check{\mathcal{D}}_{\hat{\sigma}_x+ie^{-2\alpha^{2}}\hat{\sigma}_y}\tilde{\rho}_{2}^{\rm{diss}},
\end{equation}
which shows that the $\hat{\sigma}_y$-terms, responsible for phase flip of qubits, is exponentially suppressed by $\alpha$~\cite{2019PuriP4100941009}.
Assuming that the system is initially prepared in the state
$\left|+\right\rangle$, the phase-flip probability is defined as $\mathbb{A}=\langle-|\rho|-\rangle$, where $|\pm_{1} \rangle= (|1\rangle\pm|0\rangle)/\sqrt{2}$ for the MQ, and $|\pm_{2} \rangle= (|C_\alpha^+\rangle\pm|C_\alpha^-\rangle)/\sqrt{2}$ for the MCQ.
Fig.~\ref{SMFIG3}(d) presents the temporal evolution of $\mathbb{A}_{1}$ and $\mathbb{A}_{2}$.
The comparison demonstrates that the MCQ exhibits a pronounced advantage over the MQ in suppressing phase-flip errors under the identical $\kappa$.
Figure~\ref{SMFIG3}(e) shows the phase-flip probability $\mathbb{A}_{2}$ of the MCQ under the combined influence of the parameter $\alpha$ and the evolution time. Although this probability increases for longer evolution times and smaller values of $\alpha$, $\mathbb{A}_{2}\ll1$ even at extended times when $\alpha=2$, indicating that phase-flip errors are strongly suppressed.

\section{\label{secII}Mechanical-cat-qubit gravimeter}
In this section, we present a detailed derivation of the Hamiltonian for the MCQ gravimeter. We then conduct a thorough analysis of the QFI and the sensitivity of the MCQ gravimetry scheme, including the determination of the optimal sensing time for the scheme. In addition to the sensitivity under ideal conditions, we further investigate the effects of single-phonon loss and white noise on the sensitivity.

\subsection{Hamiltonian}
A unitary transformation $\hat{U}_{0}=e^{-i\omega\hat{a}^{\dagger}\hat{a}}$ convert the Hamiltonian of the sensing system under the gravitational influence into
\begin{equation}
\hat{H}_{2}^{g}=\hat{H}_2+(mg-F)z_{0}(\hat{a}^{\dagger}e^{i\omega t}+\hat{a}e^{-i\omega t}).\label{H2g}
\end{equation}
We next project $\hat{H}_{2}^{g}$ onto the cat-state subspace spanned by $|C_\alpha^{\pm}\rangle$ using the projection operator $\hat{\mathcal{P}}_{2}$.
It is worth noting that the mean phonon number $N\simeq \alpha^2$ of the even and odd cat states $|C_\alpha^{\pm}\rangle$ can be seen as the quantum-resource number.
The projection of $\hat{H}_2$ onto the qubit subspace is a constant due to the degenerace of the two cat states.
We therefore focus on the projection of the last term.
Substituting the projected forms of the creation and annihilation operators in Eq.~\eqref{P2} into $\hat{H}_{2}^{g}$, we obtain the effective sensing Hamiltonian in the qubit subspace
\begin{equation}
\hat{\mathcal{H}}_{2}^{g} =\hbar \Omega_{2}\sigma^{x}_{2}\cos{(\omega t)},\label{H2gg}
\end{equation}
with $\Omega_{2}= 2\alpha (mg-F)z_{0}/\hbar$.
Here, terms that are exponentially suppressed in the large-$\alpha$ limit have been neglected.

\begin{figure}
\centering
\includegraphics[width=0.6\textwidth]{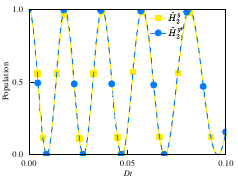}
\caption{Dynamical evolution of the system governed by the original Hamiltonian $\hat{H}_{2}^{g}$ (yellow curve) and the effective Hamiltonian $\hat{H}_{2}^{g\prime}$ (blue dotted line). Parameters used are $D=0.1\omega$, $P=3.6\omega$, and $\Omega_2=0.1\Delta_2$.}\label{SMFIG4}
\end{figure}

Figure \ref{SMFIG4} compares the dynamics generated by the original Hamiltonian $\hat{H}_{2}^{g}$ and the effective Hamiltonian $\hat{\mathcal{H}}_{2}^{g}$. The parameters are chosen as $D=0.1\omega$, and $P=3.6\omega$, and the confinement condition suppressing leakage outside the computational subspace is satisfied. 
The agreement between the two dynamics confirms the validity of the effective description.

\subsection{Quantum Fisher information and sensitivity}
The QFI of the MCQ gravimeter is evaluated  following the same procedure as that for the MQ gravimeter. 
We first solve the equation of motion, $\dot{\rho}_2(t)=-i/\hbar [\hat{\mathcal{H}}_{2}^{g},\rho_2(t)]$, for the density matrix $\rho_2(t)$ with the initial condition $\rho_{2}(0)=|C_{\alpha}^{-}\rangle\langle C_{\alpha}^{-}|$ in the decoherence-free case.

In the basis $\{|C_{\alpha}^{+}\rangle,|C_{\alpha}^{-}\rangle\}$, the density matrix is
\begin{equation}
	\rho_{2}(t)=\frac{1}{2} \begin{bmatrix}
		1-\cos\left[\frac{2\Omega_{2}}{\omega}\sin(\omega t)\right]&-i\sin\left[\frac{2\Omega_{2}}{\omega}\sin(\omega t)\right] \\
		i\sin\left[\frac{2\Omega_{2}}{\omega}\sin(\omega t)\right]&1+\cos\left[\frac{2\Omega_{2}}{\omega}\sin(\omega t)\right]
	\end{bmatrix}.
\end{equation}
Diagonalizing $\rho_{2}(t)$ yields $\rho_{2}^{\Lambda}(t)=\text{diag}(1,0)$ in the eigenbasis $\{|\lambda_{1}^{\prime}\rangle,|\lambda_{2}^{\prime}\rangle\}$, with $\lambda_{1}^{\prime}=1$ and $\lambda_{2}^{\prime}=0$.
The transformation matrix that diagonalize $\rho_{2}(t)$ is
\begin{equation}
\Lambda_{2}= \begin{bmatrix}
		i\cot\left[\frac{2\Omega_{2}}{\omega}\sin(\omega t)\right]-i\csc\left[\frac{2\Omega_{2}}{\omega}\sin(\omega t)\right]&i\cot\left[\frac{2\Omega_{2}}{\omega}\sin(\omega t)\right]+i\csc\left[\frac{2\Omega_{2}}{\omega}\sin(\omega t)\right] \\
		1&1
	\end{bmatrix}.
\end{equation}
Therefore, the derivative of the density matrix with respect to gravitational acceleration $g$ is given by
\begin{equation}
	(\partial_{g}\rho_{2})^{\Lambda}=\Lambda_{2}^{\dagger}(\partial_{g}\rho_{2})\Lambda_{2}=
	\begin{bmatrix}
		0&l_{12}^{\prime} \\
		l_{21}^{\prime}&0
	\end{bmatrix},
\end{equation}
where $l_{12}^{\prime}$ and $l_{21}^{\prime}$ denote the nonzero off-diagonal matrix elements.
According to the definition of the symmetric logarithmic derivative operator, the operator $\hat{\varsigma}_{2}$ is $\hat{\varsigma}_{2}=2\begin{bmatrix}
		0&l_{12}^{\prime} \\
		l_{21}^{\prime}&0
	\end{bmatrix}$.
Finally, with both $\hat{\varsigma}_{2}$ and $\rho_{2}^{\Lambda}(t)$, the QFI of the MCQ gravimeter is obtained as
\begin{equation}
\mathcal{F}_{2}=\frac{8Nm}{\omega^{3}\hbar}\sin^2(\omega t).
\label{F2}
\end{equation}

\begin{figure}
	\centering
	\includegraphics[width=0.6\textwidth]{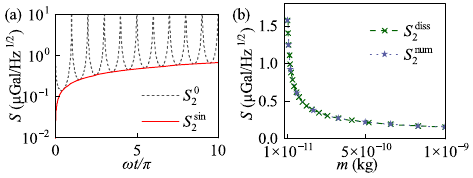}
	\caption{(a) Time evolution of the sensitivity $S_{2}^{0}$ of the MCQ gravimeter (black curve) and its envelope (red curve). Parameters: $\omega/2\pi=10~\rm{kHz}$, $D=0.1\omega$, $P=3.6\omega$, and $m=10^{-9}~\rm{kg}$.  (b) Sensitivities of the MCQ gravimeter at the optimal sensing time $t_{2}=\pi/(2\omega)$ as a function of the particle mass $m$, under white thermal  noise.  $S_{2}^{\text{diss}}$  and $S_{2}^{\text{num}}$ denote the analysis result and numerical result of sensitivity, respectively. The same parameters as in panel (a) are adopted. Other parameters are $Q=10^{8}$,  $T=10~\rm{mK}$ and $\Omega_2=\pi\omega/4$.}
	\label{SMFIG5}
\end{figure}

Since the MCQ gravimetry also employs the Rabi measurement, its sensitivity analysis follows the same procedure as for the MQ scheme.
In our MCQ gravity sensing scheme, $\hat{O}_2=|C_{\alpha}^{+}\rangle\langle C_{\alpha}^{+}|$ is the even-cat-state
occupation operator, and the initial state is $|\psi_{2}(0)\rangle=|C_{\alpha}^{-}\rangle$. Under ideal conditions, the system evolves according to the time evolution operator $\hat{U}_{2}=\exp\left[-\frac{i}{\hbar}\int_{0}^{t}\hat{\mathcal{H} }_{2}^{g}(t^{\prime})dt^{\prime}\right]=\exp\left[-\frac{i\Omega_{2}}{\hbar\omega}\sin(\omega t)\right]$. 
The evolved state is
\begin{equation}
	|\psi_{2}(t)\rangle=\cos\left[\frac{\Omega_{2}}{\omega}\sin(\omega t)\right]|C_{\alpha}^{-}\rangle-i\sin\left[\frac{\Omega_{2}}{\omega}\sin(\omega t)\right]|C_{\alpha}^{+}\rangle.
\end{equation}
The corresponding mean value of the even-cat-state occupation operator $\hat{O}_2$ is therefore $\bar O_{2}=\sin^{2}\left[\frac{\Omega_{2}}{\omega}\sin(\omega t)\right]$.
Its variance and derivative with respect to $g$ are $\delta O_{2}=\frac{1}{2}|\sin[\frac{\Omega_{2}}{\omega}\sin(\omega t)]|$ and $
	\frac{\partial \bar O_{2}}{\partial g}=\frac{2\sqrt{N} z_{0}m}{\hbar\omega}\sin(\omega t)\sin\left[\frac{2\Omega_{2}}{\omega}\sin(\omega t)\right]$. 
The corresponding sensitivity is
\begin{equation}
S_{2}^{0}=\sqrt{\frac{\hbar\omega}{8Nm}}\frac{\omega\sqrt{t}}{|\sin(\omega t)|}.\label{S02}
\end{equation}
The optimal sensing time is determined by analyzing the time dependence of the sensitivity $	S_{2}^{0}$.
As shown in Fig.~\ref{SMFIG5}, the black dashed curve represents  $	S_{2}^{0}$ as a function of the sensing time $t$. 
Due to the oscillatory factor $|\sin(\omega t)|$ in the denominator, the sensitivity exhibits a sequence of local minima occurring at $\sin(\omega t)=1$, i.e., $\omega t=(2n+1)/2$.
Under this condition, the sensitivity reduces to the envelope function $S_{2}^{\text{sin}}=\omega\sqrt{\frac{\hbar\omega t}{8Nm}}$.
This envelope is shown as the red solid curve in Fig.~\ref{SMFIG5}(a) and clearly bounds the oscillatory behavior of the black dashed curve.
The first minimum occurs at $n=0$.
The optimal sensing time for the MCQ gravity measurement scheme is therefore $t_{2}=\pi/(2\omega)$.
At this optimal sensing time, the sensitivity of the MCQ gravimetry scheme becomes
\begin{equation}
S_{2}=\omega\sqrt{\frac{\hbar\pi}{16Nm}},\label{S2}
\end{equation}
and the corresponding QFI reaches its maximal value
\begin{equation}
\mathcal{F}_{2}^{g}=\frac{8Nm}{\hbar\omega^{3}}.\label{F2g}
\end{equation}
Equations~\ref{S2} and~\ref{F2g} represent that the sensitivity $S_{2}$ matches the QFI $\mathcal{F}_{2}^{g}$, demonstrating that the proposed measurement protocol is optimal under ideal conditions. Most importantly, compared to the MQ gravimeter whose sensitivity is $S_{1}$ in Eq.~\eqref{S1}, the MCQ gravimetry scheme not only fully exploits the mass enhancement provided by the levitated particles, but also achieves an additional enhancement factor of $\sqrt{2N}$.

Under the white thermal noise, the system dynamics is governed by the master equation
\begin{equation}
		\dot{\rho}_{2}'=-\frac{i}{\hbar}[\hat{H}_{2}^{g},\rho_{2}']+\kappa[(n_{\rm{th}}+1)\check{\mathcal{D}}_{\hat{a}}+n_{\rm{th}}\check{\mathcal{D}}_{\hat{a}^{\dagger}}]\rho_{2}',
\end{equation}
where $\rho_{2}'$ denotes the density matrix under the white thermal noise and $n_{\rm{th}}=1/\{\exp[\hbar\omega/(k_{B}T)]-1\}$ is the thermal phonon occupation number at temperature $T$, with $k_{B}$ the Boltzmann constant. 
Projecting the master equation onto the cat-state subspace using Eq.~\eqref{P2} yields
\begin{equation}
\dot{\rho}_{2}^{\text{diss}}=-\frac{i}{\hbar}[\hat{\mathcal{H}}_{2}^{g},\rho_{2}^{\text{diss}}]+\kappa N[(n_{\rm{th}}+1)\check{\mathcal{D}}_{\hat{\sigma}_{2}^{x}+ie^{-2N}\hat{\sigma}_{2}^{y}}+n_{\rm{th}}\check{\mathcal{D}}_{\hat{\sigma}_{2}^{x}-ie^{-2N}\hat{\sigma}_{2}^{y}}]\rho_{2}^{\text{diss}},
\end{equation}
with $\rho_{2}^{\text{diss}}$ denotes the density matrix projected onto the cat-state subspace in the presence of white thermal noise.
In the large-$\alpha$ limit, the terms proportional to $e^{-2N}$ are exponentially suppressed and are neglected.
The master equation then reduces to
\begin{equation}
	\dot{\rho}_{2}^{\text{diss}}=-\frac{i}{\hbar}[\hat{\mathcal{H}}_{2}^{g},\rho_{2}^{\text{diss}}]+\kappa N(2n_{\rm{th}}+1)\check{\mathcal{D}}_{\hat{\sigma}_{2}^{x}}\rho_{2}^{\text{diss}}.\label{rho}
\end{equation}

\begin{table*}
	\caption{\label{SimulationPara1}
		Sensitivity comparison between our gravimetry scheme and other levitated-particle-based gravimetry schemes~\cite{2013ScalaP180403180403,2025WangP120803120803}, together with the corresponding parameters.
	}
	\begin{ruledtabular}
		\begin{tabular}{lcccccc}
			\makecell{Parameters}&
			\makecell{\textrm{Frequency}\\$\omega/2\pi~(\mathrm{kHz})$}&
			\makecell{\textrm{Mass}\\$m~(\mathrm{kg})$}&
			\makecell{\textrm{Duffing strength}\\$D/2\pi~(\mathrm{kHz})$}&
			\makecell{\textrm{Pumping}\\$P/2\pi~(\mathrm{kHz})$}&
			\textrm{Sensing time}&
			\makecell{\textrm{Sensitivity}\\$S~(\mu\mathrm{Gal}/\sqrt{\mathrm{Hz}})$}\\
			\colrule
			\makecell{Nanoparticle gravimeter\\in ideal case~\cite{2013ScalaP180403180403}} & 10 & Mass-independent & 0 & 0 & $t_S=2~\mathrm{ms}$ & $S_S=501.80$ \\
            \colrule
			\makecell{Free falling Nanoparticle\\gravimeter for single partial\\ in ideal case\cite{2025WangP120803120803}}& 10 & Mass-independent & 0 & 0 & $2t_W=2~\mathrm{ms}$ & $S_W=17$ \\
            \colrule
			\makecell{MQ gravimeter\\in ideal case} & 10 & $10^{-9}$& 1 & 36 & $t_1=\pi/\omega$ & $S_1=1.28$\\
            \colrule
			\makecell{MQ gravimeter\\with thermal noise\\with $Q=10^8$ and $T=10~\rm{mK}$} & 10 & $10^{-9}$& 1 & 36 & $t_1=\pi/\omega$& $S_{1}^{\rm{diss}}=1.52$\\
            \colrule
			\makecell{MCQ gravimeter\\in ideal case} & 10 & $10^{-9}$& 1 & 36 & $t_2=\pi/(2\omega)$ & $S_2=0.15$\\
            \colrule
			\makecell{MCQ gravimeter\\with thermal noise\\with $Q=10^8$ and $T=10~\rm{mK}$} & 10 & $10^{-9}$& 1 & 36 & $t_2=\pi/(2\omega)$ & $S_{2}^{\rm{diss}}=0.16$\end{tabular}
	\end{ruledtabular}
	\label{table}
\end{table*}

Solving Eq.~\eqref{rho} for the initial state $|\psi_{2}(0)\rangle=|C_{\alpha}^{-}\rangle$ yields the mean value of $\hat{O}_2$ as
\begin{equation}
	\bar{O}_{2}^{\text{diss}}=\frac{1}{2}-\frac{1}{2}\cos\left[\frac{2\Omega_{2}}{\omega}\sin(\omega t)\right]e^{-2\kappa N(2n_{\rm{th}}+1)t}.
\end{equation}
Its derivative with respect to the gravitational acceleration $g$ is 
\begin{equation}
	\frac{\partial \bar{O}_{2}^{\text{diss}}}{\partial g}=\frac{2\sqrt{N} z_{0}m}{\hbar\omega}\sin(\omega t)\sin\left[\frac{2\Omega_{2}}{\omega}\sin(\omega t)\right]e^{-2\kappa N(2n_{\rm{th}}+1)t}.
	\label{partial}
\end{equation}
Thus, the corresponding standard deviation is
\begin{equation}
\delta O_{2}^{\text{diss}}=\frac{1}{2}\sqrt{[1-e^{-4\kappa N(2n_{\rm{th}}+1)t}]\cos^{2}\left[\frac{2\Omega_{2}}{\omega}\sin(\omega t)\right]+\sin^{2}\left[\frac{2\Omega_{2}}{\omega}\sin(\omega t)\right]}.
	\label{p2diss}
\end{equation}
Substituting Eqs.~\eqref{partial} and~\eqref{p2diss} into the sensitivity yields 
\begin{equation}
S_{2}^{\text{diss}}=\sqrt{\frac{\hbar\pi\omega^{3}}{16Nm}}e^{2\kappa N(2n_{\rm{th}}+1)t_{2}},
\end{equation}
at the optimal sensing time $t_{2}=\pi/(2\omega)$ and $\Omega_2=\pi\omega/4$.

We also compute the sensitivity under thermal white noise by directly solving the master equation given in Eq.~\eqref{rho} and compare the numerical results $S_{2}^{\text{diss}}$ with the analytical solution $S_{2}^{\text{num}}$ as functions of $m$ in Fig.~\ref{SMFIG5}(b).
As shown in Fig.~\ref{SMFIG5}(b), $S_{2}^{\text{diss}}$ and $S_{2}^{\text{num}}$ are in perfect agreement, confirming  that $S_{2}^{\text{diss}}$ correctly describes the sensitivity of the MCQ gravimeter under thermal white noise, for the chosen parameters.

We finally present the sensitivities  of our approach under different conditions in Table~\ref{table}, along with a comparison to the sensitivities previously reported  gravity sensing approaches based on levitated particles under  ideal conditions from Refs.~\cite{2013ScalaP180403180403,2025WangP120803120803}.
As shown in Table~\ref{table}, our MQ gravimeter achieves the best sensitivity among existing levitated-particle gravimeters in the single-partical case without requiring free-falling operation, which greatly reduces the device size and enables miniaturization and integration. Moreover, the MCQ gravimeter further improves the sensitivity by two orders of magnitude while remaining only weakly affected by dissipation.

\section{\label{secIII}Dissipation sources}
In the analysis of the sensitivity of the MCQ gravimeter constructed using the CM mode of levitated mesoscopic particles under dissipative conditions, we adopt a quality factor of $Q=10^8$ for the levitated particle.
In this section, we consider two primary sources of dissipation, i.e., the gas damping and the blackbody radiation. We analyze the resulting dissipation of the particle's CM mode and demonstrate that a quality factor of $Q=10^8$ is achievable.

\subsection{Gas damping}
For the levitated particle, the gas damping is the dominant source of decoherence of the mechanical mode.
Under high vacuum conditions, the dissipation $\kappa_\text{gas}/2\pi$ caused by collisions with residual gas molecules~\cite{2023HeiP7360273602,2012GieselerP103603103603} is described by
\begin{equation}
\frac{\kappa_{\text{gas}}}{2\pi}=3\eta\frac{r}{m}\frac{0.619}{0.617+K_{n}}(1+c_{K}),
\end{equation}
where $c_{K}=0.31K_{n}/(0.785+1.152K_{n}+K^{2}_{n})$ is a positive function of the Knudsen number $K_{n}=\frac{\bar{L}}{r}$, $\bar{L}$ is the mean free path of the molecules, and $r$ is the radius of the levitated particle.
The relationship between $\bar{L}$ and the viscosity coefficient $\eta$ is $\eta=n_{\text{gas}}\bar{L}\sqrt{{2m_{\text{gas}}k_{B}T}/{\pi}}$,
where $m_{\text{gas}}$ and $n_{\text{gas}}$ denote the mass and number density of the gas molecules, respectively.
Using the gas pressure relation $\mathfrak{P}=n_{\text{gas}}k_{B}T$, we obtain
\begin{equation}
\eta=\mathfrak{P}l\sqrt{\frac{2m_{\text{gas}}}{\pi k_{B}T}}.	\label{eta}
\end{equation}
Under the high-vacuum conditions, the mean free path of gas molecules is much larger than the particle radius, i.e., $\bar{L}\gg r$. Thus, the Knudsen number $K_{n}$ is very large. 
Therefore, we have $c_{K}\simeq0$ and $\frac{\kappa_{\text{gas}}}{2\pi}\simeq 3\eta\frac{r}{m}\frac{0.619}{K_{n}}$.
The substitution of Eq.~\eqref{eta} into $\kappa_{\text{gas}}$ yields
\begin{equation}
\frac{\kappa_{\text{gas}}}{2\pi}=\frac{0.571\mathfrak{P}}{(m\varrho^{2})^{1/3}}\sqrt{\frac{m_{\text{gas}}}{k_{B}T}},
\end{equation}where $\varrho=3m/(4\pi r^3)$ is the density of the levitated diamond particle. 

\begin{figure}
\centering
\includegraphics[width=0.6\textwidth]{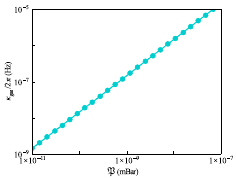}
\caption{Gas damping rate  $\kappa_{\text{gas}}$ versus the pressure $\mathfrak{P}$. The parameters used in the calculation are  $m=10^{-9}~\rm{kg}$, $\rho=3500~\rm{kg/m^{3}}$, $T=10~\rm{mK}$, and $m_{\text{gas}}=28~\rm{aum}$.}\label{SMFIG6}
\end{figure}

Figure~\ref{SMFIG6} shows $\kappa_{\text{gas}}/2\pi$ for a levitated diamond particle with $\varrho=3500~\rm{kg/m^{3}}$ and $m=10^{-9}~\rm{kg}$ in molecules as a function of the air pressure at temperature $T=10~\rm{mK}$.
The mass of the gas molecules is taken to be $m_{\text{gas}}=28~\rm{aum}$.
The high-vacuum conditions assumed in the dissipation calculations are experimentally accessible~\cite{PhysRevLett.132.133602}.

\subsection{\label{secIIIB}Blackbody radiation}
The coupling between the particle and the blackbody radiation field leads to momentum transfer from photons to the particle, thereby inducing a damping force (dissipation) and a random force (fluctuation)  acting on the particle's CM motion.
For the CM mode of the levitated particle, the Langevin equation for the momentum operator is
\begin{equation}
	\frac{d\hat{p}}{dt}=-m\omega^{2}\hat{z}-\kappa_{\rm{b}}\hat{p}+\hat{F}(t),
	\label{Langevin}
\end{equation}
where $\hat{p}$ and $\hat{z}$ are the momentum and position operators of the CM motion, satisfying the commutation relation $[\hat{z},\hat{p}]=i\hbar$.
Here $\hat{F}(t)$ is the fluctuation force operator, and $\kappa_{\rm{b}}$ is the dissipation rate.
The quantum fluctuation-dissipation theorem~\cite{2001PottierP327344},
\begin{equation}
	C_{FF}(\omega^{\prime})=\hbar\coth{\left(\frac{\hbar\omega^{\prime}}{2k_{B}T}\right)}\xi_{FF}(\omega^{\prime}),
	\label{FD}
\end{equation}
relates the fluctuation force operator to the symmetrized equilibrium correlation function $C_{FF}(\omega^{\prime})$ in the frequency domain and to the dissipative component $\xi_{FF}(\omega^{\prime})$ of the response function $\chi_{FF}(\omega^{\prime})$. In our system, the symmetrized equilibrium correlation function is $C_{FF}(\omega^{\prime})=\frac{1}{2}\int_{-\infty}^{\infty}\langle\{\hat{F}(t),\hat{F}(0)\}\rangle e^{i\omega^{\prime}t}dt$, which reduces to
\begin{equation}
C_{FF}(0)=\frac{1}{2}\int_{-\infty}^{\infty}\langle\{\hat{F}(t),\hat{F}(0)\}\rangle dt\label{CFF}
\end{equation}in the limit $\omega^{\prime}\rightarrow0$. 
Under the Markov approximation, the force operator satisfies $\langle\hat{F}(t)\hat{F}(t^{\prime})\rangle=2\mathbb{D}\delta(t-t^{\prime})$, where $\mathbb{D}$ is the diffusion coefficient. 
Thus, Eq.~\eqref{CFF} reduces to 
\begin{equation}
    C_{FF}(0)=2\mathbb{D}.\label{dcff0}
\end{equation}
The response function $\chi_{FF}(\omega^{\prime})$ can be obtained from the equation of motion, $m\ddot{z}+m\kappa_{\rm{b}}\dot{z}+m\omega^{2}z=F(t)$.
Making the Fourier transform to it, we obtain
\begin{equation}
-m\omega^{\prime2}\int z(t)e^{i\omega^{\prime}t}dt-im\kappa_{\rm{b}}\omega^{\prime}\int z(t)e^{i\omega^{\prime}t}dt+m\omega^{2}\int z(t)e^{i\omega^{\prime}t}dt=F(\omega^{\prime}).
\end{equation}
This gives response function associated with the force $\chi_{FF}(\omega^{\prime})\equiv\frac{F(\omega^{\prime})}{\int z(t)e^{i\omega^{\prime}t}dt}=m(-\omega^{\prime2}-i\kappa_{\rm{b}}\omega^{\prime}+\omega^{2})$.
Its imaginary part gives $\xi_{FF}(\omega^{\prime})=m\kappa_{\rm{b}}\omega^{\prime2}$.
Substituting $\xi_{FF}(\omega^{\prime})$ into Eq.~\eqref{FD} results in
\begin{equation}
C_{FF}(\omega^{\prime})=m\kappa_{\rm{b}}\hbar\omega^{\prime}\coth{\left(\frac{\hbar\omega^{\prime}}{2k_{B}T}\right)},
\end{equation}
which expresses the relationship between fluctuations induced by blackbody radiation and the corresponding dissipation rate.
In the limit $\omega^{\prime}\rightarrow0$, neglecting the higher-order terms, we obtain $\coth{\left[\hbar\omega^{\prime}/(2k_{B}T)\right]}\approx2k_{B}T/(\hbar\omega^{\prime})$, which leads to
\begin{equation}
	2\mathbb{D}=2m\kappa_{\rm{b}}k_{B}T,\label{dd1}
\end{equation}
where Eq. \eqref{dcff0} has been used.
For a dielectric sphere of radius $r$ with polarizability $\beta(\omega^{\prime})=\left(\frac{\epsilon-1}{\epsilon+2}\right)r^{3}$ in blackbody radiation, and assuming that the frequency dependence of the permittivity $\epsilon$ can be neglected, the diffusion coefficient is~\cite{2022SinhaP204002204002}
\begin{equation}
	2\mathbb{D}=\frac{d\langle\Delta p^{2}\rangle }{dt}=\hbar^{2}\frac{1024\pi^{2}}{135}r^{6}c\left|\frac{\epsilon-1}{\epsilon+2}\right|^{2}\left(\frac{k_{B}T}{\hbar c}\right)^{9},\label{dd2}
\end{equation}
where $c$ is the speed of light in vacuum. The combination of Eqs. \eqref{dd1} and \eqref{dd2} gives the form of $\kappa_\text{b}$.

For a levitated diamond particle with mass $m=10^{-9}~\text{kg}$ and density $\varrho=3500~\rm{kg/m^{3}}$, using a dielectric constant $\epsilon=5.7$, the resulting dissipation rate due to blackbody radiation is on the order of $\kappa_{\rm{b}}\sim10^{-39}~\text{Hz}$.
It is many orders of magnitude smaller than the dissipation rate $\kappa_{\rm{gas}}$ caused by collisions with residual gas molecules and can therefore be safely neglected. 
Hence we take $\kappa=\kappa_{\rm{gas}}+\kappa_{\rm{b}}\approx\kappa_{\rm{gas}}$. 
The CM quality factor is determined by
\begin{equation}
	Q=\frac{\omega}{\kappa_{\rm{gas}}}.
\end{equation}
In our scheme, the minimum oscillation frequency is $\omega/2\pi=10~\rm{kHz}$. 
Under experimentally achievable high-vacuum conditions, the maximum dissipation rate shown in Fig.~\ref{SMFIG6} is $\kappa_{\rm{gas}}/2\pi=10^{-5}~\rm{Hz}$, which fully satisfies $Q\ge10^{9}$.
Therefore, the quality factor used in the main text is experimentally feasible.

\bibliography{SMReference}